\documentstyle[11pt]{article}

\advance\voffset by -1.8cm
\advance\hoffset by -1.9cm
\def\be{\begin{equation}}
\def\ee{\end{equation}}
\def\Cop{\bbbc}
\def\Zop{\bbbz}

\def\Nop{\bbbn}
\def\bbbz {{\sf Z\!\!Z}}

\def\bbbn {{\rm I\!N}}
\def\bbbc{{\mathchoice {\setbox0=\hbox{$\displaystyle\rm C$}\hbox{\hbox
to0pt{\kern0.4\wd0\vrule height0.9\ht0\hss}\box0}}
{\setbox0=\hbox{$\textstyle\rm C$}\hbox{\hbox
to0pt{\kern0.4\wd0\vrule height0.9\ht0\hss}\box0}}
{\setbox0=\hbox{$\scriptstyle\rm C$}\hbox{\hbox
to0pt{\kern0.4\wd0\vrule height0.9\ht0\hss}\box0}}
{\setbox0=\hbox{$\scriptscriptstyle\rm C$}\hbox{\hbox
to0pt{\kern0.4\wd0\vrule height0.9\ht0\hss}\box0}}}}
\def\empty{{\emptyset}}

\def\pmb#1{\setbox0=\hbox{#1}%
 \kern-.025em\copy0\kern-\wd0
 \kern.05em\copy0\kern-\wd0
 \kern-.025em\raise.0433em\box0 }
\def\sq{\hbox{\rlap{$\sqcap$}$\sqcup$}}
\def\qed{\ifmmode\sq\else{\unskip\nobreak\hfil
\penalty50\hskip1em\null\nobreak\hfil\sq
\parfillskip=0pt\finalhyphendemerits=0\endgraf}\fi}

\def\A{{\cal A}}
\def\a{{\bf a}}
\def\bOmega{{\bf \Omega}}
\def\hOmega{\widehat{\Omega}}

\def\H{{\cal H}}

\def\M{{\cal M}}

\def\G{{\cal G}}

\def\o{{\otimes}}

\def\bbbone {{\mathchoice {{\rm 1\mskip-4mu l}} {{\rm 1\mskip-4mu l}}
{{\rm 1\mskip-4.5mu l}} {{\rm 1\mskip-5mu l}}}}

\begin{document}

\thispagestyle{empty}
\def\thefootnote{\fnsymbol{footnote}}
\begin{flushright}
  HUTP-97/A030 \\
  MIT-CTP-2655 \\
  hep-th/9707051
\end{flushright} \vskip 2.0cm

\begin{center}\LARGE
{\bf Tensor Constructions of Open  String Theories II: 
Vector bundles, D-branes and orientifold groups}
\end{center} \vskip 1.0cm
\begin{center}\large
       Matthias R.\ Gaberdiel%
       \footnote{E-mail  address: {\tt gaberd@string.harvard.edu}}
       and 
       Barton Zwiebach%
       \footnote{E-mail  address: {\tt zwiebach@irene.mit.edu}}
       \end{center}
\vskip0.5cm

\begin{center}
Lyman Laboratory of Physics\\
Harvard University\\
Cambridge, MA 02138
\end{center}
\vskip 4em
\begin{center}
July 1997
\end{center}
\vskip 1cm
\begin{abstract}

A generalized Chan-Paton construction is presented which is analogous to
the tensor product of vector bundles. To this end open string theories are
considered where the space of states decomposes into sectors whose product
is described by a semigroup. The cyclicity properties of the open
string theory are used to prove that the relevant semigroups are direct
unions of Brandt semigroups. The known classification of Brandt semigroups
then implies that all such theories have the structure of a theory with
Dirichlet-branes. 
We also describe the structure of an
arbitrary orientifold group, and show that the truncation to the
invariant subspace defines a consistent open string theory.
Finally, we analyze the possible orientifold projections of a theory with
several kinds of branes.

\end{abstract}

\vfill

\setcounter{footnote}{0}
\def\thefootnote{\arabic{footnote}}
\newpage

\renewcommand{\theequation}{\thesection.\arabic{equation}}

\section{Introduction}
\setcounter{equation}{0}

It is well known that the algebraic structure of open string theory is
rather rich. For example, it is possible to attach Chan-Paton
indices to open string theories \cite{chanpaton} without destroying the
structure of the underlying theory,
and, as has become apparent recently, more elaborate constructions
involving Dirichlet-branes are also possible
\cite{PS,BSag,GimPol,BG}. The traditional Chan-Paton 
constructions were classified some time ago using Wedderburn's theorem
\cite{MarSag}; in this paper we show that the new constructions can
be understood as arising from tensor products of vector bundles over
semigroups, and we classify the relevant semigroups completely.
\smallskip

In a recent work \cite{GZ1}, we reexamined the algebraic structure of open
string theory using the framework of open string field theory
\cite{Witten,Thorn,Z2}. We showed that a classical string field theory can
be constructed provided that the space of states has the structure of a 
homotopy associative star algebra $\A$ \cite{Stasheff} with an odd
invertible bilinear form. This structure is preserved if we tensor to $\A$
a homotopy associative star algebra $\a$ with an even form. In the case
where $\a$ does not have any spacetime interpretation, the positivity of
the string field action requires that the even form is positive
definite on the cohomology of $\a$. 
This restriction is strong enough to prove that 
all such constructions are physically equivalent to the familiar
Chan-Paton constructions, where $\a$ is a direct sum of full matrix
algebras. Unoriented string theories are then obtained as the restriction of
the tensor space $\A\o\a$ to the subspace which is invariant under the 
action of the twist operator $\bOmega=\Omega\o\omega$.  
\medskip

In this paper we generalize this construction in two directions, both of
which use additional structure of $\A$. On the one hand, we consider
general orientifold group projections of a tensor theory $\A\o\a$, and on
the other hand, we modify the construction of a tensor theory itself. All
of our considerations will be at the level of classical open string
theory. 
\smallskip

Orientifold projections are possible if $\A$ has a symmetry which is
described by a compact orientifold group $\G$. We demonstrate that $\G$ is of
the form $\G=G\cup \hOmega G$, where $G$ is the subgroup of orientation
preserving symmetries, and $\hOmega\in\G$ reverses the orientation. We then
show how $\A$ can be truncated consistently to the invariant subspace under
the action of $\G$. This truncation can also be applied to the tensor theory
$\A\o\a$, whose orientifold group is $(G\times H) \cup (\hOmega G \times
\hat\omega H)$ if $\A$ and $\a$ have the orientifold groups $\G=G\cup
\hOmega G$ and $\H=H\cup\hat\omega H$, respectively.  In this way we can
obtain the non-supersymmetric theories with gauge groups $SO(32-n)\times
SO(n) \times SO(32-n)\times SO(n)$ \cite{BSag}.\footnote{This is a
convenient description, but as explained in \cite{BG}, these theories can
also be obtained using D-branes, and therefore can be described using the
fiberwise tensor product to be described below.}
\smallskip

To describe more general tensor products we consider theories which have
the structure of a vector bundle over a semigroup $S$. (A semigroup
is a set with an associative product.) This means that $\A$
decomposes into sectors $\A_s$ which are labeled by elements $s\in S$, and
that the product of the sectors is described by the product in $S$. If both
$\A$ and $\a$ have this bundle structure, we consider the {\it fiberwise}
tensor product $\A\o_S\a$ which again has the structure of a vector bundle
over $S$, and we show that it defines a consistent string field theory.

We are able to prove that the semigroup $S$ that arises as the base space
of an open string theory has to be a direct union of Brandt semigroups and
trivial semigroups. Brandt semigroups have been classified completely, and
they are characterized in terms of a group $G$ and a positive
integer $n$. In terms of string theory, the Brandt semigroup characterized
by $(G,n)$ arises in a configuration with $n$ different types of D-branes,
where, in addition, each open string sector decomposes into subsectors
labeled by elements in $G$.

We also analyze the possible twist truncations of tensor bundles,
and we give the most general solution for the case where $\Omega^4=+1$.
This analysis illuminates and generalizes that of Ref.\cite{GimPol}, where
the combined system of D5-branes and D9-branes in type I string theory was
considered.

\section{A brief review of the tensor construction}
\setcounter{equation}{0}

Let us review briefly some aspects of the tensor construction by means of
which an algebra $\a$ can be attached to an open string theory
$\A$. (More details can be found in \cite{GZ1}.) We shall mainly
consider open string theories which possess the {\it basic algebraic
structure} 
\begin{list}{(\roman{enumi})}{\usecounter{enumi}}
\item $\A$ is a homotopy-associative star algebra.
\item $\A$ is equipped with an invertible odd bilinear form  
$\langle\cdot,\cdot \rangle$ with cyclicity properties.
\item The sesquilinear form  $(A, B) = \langle A^*, c_0 B\rangle$ 
is positive definite
on physical states. 
\end{list}
$\A$ is a homotopy-associative algebra if there exist multilinear maps
$b_n: \A^{\otimes n} \to \A$, $n\geq1$, which satisfy a number of
consistency conditions. Here $b_1$ is the BRST-operator $Q$, the map
$b_2$ defines the string product, and $b_n$ for $n\geq 3$ represent higher
products. The consistency conditions imply, for example, that $Q^2=0$, and
that $Q$ is a derivation of the string product. $\A$ is a star algebra if
there exists an involution $*$ on $\A$ so that 
$(b_n (A_1 , \ldots , A_n))^* = \pm b_n (A_n^*, \ldots , A_1^*)$,
where $\pm$ is a sign factor that will not be relevant here.
The star conjugation enables one to impose a reality condition on
the string field thus ensuring the reality of the action. 
The bilinear form in (ii) satisfies 
$\langle A , B\rangle = \pm \langle B , A \rangle$,
and the cyclicity conditions take the form 
$\langle A_1 , b _{n}(A_2 , \ldots , A_{n+1} ) \rangle = 
\pm \langle A_2, b _{n} \, (A_3, \ldots , A_{n+1}, A_1  )\rangle$, 
which make the bilinear form structurally similar to a trace on the
algebra. In addition, 
$\overline {\langle A , B\rangle} = \langle B^*, A^* \rangle.$
Finally, the positive definite form in (iii) guarantees that physical
states have positive norms.
\medskip

An open string theory is  {\it twist invariant} if there exists
a (linear) twist operator $\Omega$ satisfying
\begin{itemize}
\item[(a)] $\Omega b_n ( A_1 , \ldots , A_n ) = 
\pm b_n ( \Omega A_n ,\ldots , \Omega A_1 )$.
\item[(b)] $\Omega$ commutes with the star-conjugation.
\item[(c)] $\langle \Omega(A) , \Omega(B) \rangle = \langle A , B \rangle$. 
\end{itemize}
The operator $\Omega^2$  is a (not necessarily trivial) automorphism
of the homotopy-associative algebra. In a twist invariant theory, we can
restrict the theory to the eigenspace of $\Omega$ of eigenvalue $+1$,
thereby preserving the classical master equation. 
\medskip

The algebra $\a$ is a homotopy associative star algebra equipped with an
{\it even} bilinear form, but otherwise satisfying  the exact analogs of
(i) and (ii). This is sufficient to show that the tensor theory
$\A\otimes\a$ can be given the structure of a homotopy associative star
algebra with an odd bilinear form that satisfies conditions (i) and (ii). 
If $\a$ has no spacetime interpretation (as we will assume here), condition
(iii) requires that the bilinear form on $\a$ is positive definite. It can
then be shown that every such tensor theory is physically equivalent to
one, where $\a$ is a direct sum of full matrix algebras
$\a = \oplus_i \M_{n_i} (\Cop)$. In this case the bilinear form is 
$\langle a, b\rangle = \hbox{Tr} (ab)$, and star conjugation is 
hermitian conjugation. 
\smallskip

Every matrix algebra $\M_{n}(\Cop)$ possesses twist operators $\omega$
which satisfy the analogs of (a)--(c) above. If $\A$ is twist invariant
with twist operator $\Omega$, then the tensor theory $\A\otimes \M_n(\Cop)$
is twist invariant, where $\bOmega=\Omega\otimes \omega$. Since all 
(linear) automorphisms of $\M_n(\Cop)$ are inner, every twist operator
$\omega$ is of the form $\omega(a) = J a^t J^{-1}$, where
$J\in\M_n(\Cop)$. If we demand that the automorphism $\omega^2$ is equal to
the identity (as in \cite{GZ1} but not necessarily here), it follows that
there are two inequivalent choices $J_\pm$ for the matrix 
$J$, $J_+ = \bbbone$ and $J_- = \pmatrix{0 & \bbbone\cr -\bbbone & 0 }$,
where we have assumed in the second case that the matrix algebra is of even
rank.

\section{Symmetry groups and projections}
\setcounter{equation}{0}

As mentioned in the previous section, it is possible to truncate a twist
invariant theory to the subspace of twist invariant states.  While the
resulting theory has an action that satisfies the classical master
equation, the vector space of twist invariant states typically does not
have the structure of a homotopy associative algebra any more.  In this
section we shall 
first consider truncations to the invariant subspace of orientation
preserving symmetry groups; these constructions always lead to string
theories whose underlying vector space possesses the structure of an
homotopy associative algebra, and the classical master equation is
therefore manifestly satisfied. We shall then explain how the general case
of orientifold groups can be treated.
\smallskip

Suppose that a string theory has a (orientation preserving) symmetry 
described by a compact group $G$. 
This means that
\begin{list}{(\roman{enumi})}{\usecounter{enumi}}
\item $G$ acts on $\A$, {\it i.e.} every $g\in G$ defines a linear map
$g:\A \rightarrow \A$, and $g\circ h = (gh)$.
\item The bilinear form  is invariant under $G$, {\it i.e.}
$\langle g(A), g(B) \rangle = \langle A, B \rangle$ for
every $g\in G$. 
\item The action of $G$ commutes with the products, 
{\it i.e.} $g\, b_n(A_1, \ldots, A_n) = b_n(g A_1, \ldots , g A_n)$ for
every $g\in G$. 
\item There exists an involution on $G$, $g\mapsto g^\ast$, so that
$(g A)^\ast = g^\ast A^\ast$ for all $g\in G$.
\end{list}  

We can then truncate the string field theory to the invariant subspace 
\be
\A^G := \{ A\in\A \;|\; g(A)=A \,,\quad \hbox{for all $g\in G$} \} \,.
\ee
First of all, $\A^G$ inherits the structure of a homotopy-associative
algebra, as (iii) implies that $b_n(A_1,\ldots , A_n) \in \A^G$
for $A_i\in\A^G$.  Similarly, $\A^G$ has a star structure because of (iv). 

Next, as $G$ is compact, we may decompose $\A$ as 
$\A = \bigoplus_{R} \A_R$, where $R$ labels the irreducible representations
of $G$, and the subspace $\A_R$ transforms as $R$ under the action of $G$
\cite{BtD}. (In particular, $\A_{1}=\A^G$, where $R=1$ is the trivial
representation.)  It then follows from (ii) that $\A^G$ is
orthogonal to $\A_R$ for ${R\ne 1}$ \cite{BtD}. 
This implies, in particular, that the bilinear form
restricted to $\A^G$ is invertible. It is then manifest that $\A^G$ has the
basic algebraic structure which guarantees that the associated master
action satisfies the classical master equation \cite{GZ1}.
\smallskip

This analysis can be generalized to the case where the symmetry is an
{\it orientifold} group $\G$, {\it i.e.} a group which also contains elements
that reverse the orientation of the string. Mathematically speaking this
means that there exists a group homomorphism 
$\sigma: \G\rightarrow \{1,-1\}\simeq \Zop_2$ (where $\sigma(g)=1$ for
order preserving group elements, and  $\sigma(h)=-1$ otherwise), and that 
(iii) is modified to
\be
\hbox{(iii')}\qquad \qquad
g\, b_n(A_1, \ldots, A_n) = (-1)^{ {(1-\sigma(g))\over 2} e_n(A_i) }\;
b_n(g A_{\tau_{\sigma (g)}(1)}, \ldots , g A_{\tau_{\sigma (g)}(n)}) \,,
\ee
where $g\in\G$, $e_n(A_i)$ is the sign-factor of \cite[(5.6)]{GZ1}, 
$\tau_1=\mbox{id}$, and $\tau_{-1}:(1,2,\ldots,n)\mapsto(n,\ldots,2,1)$.
The structure of the orientifold group is then described by\footnote{This
refines a similar discussion in Ref.\cite{GimPol}.}
\smallskip

{\bf Proposition:} If $\G$ contains an element $\hOmega$ which reverses the
order, then $\G$ has the structure $\G=G\cup\hOmega G$, where $G$ is the 
subgroup of elements that preserve the order. Moreover, 
$\hOmega G =G\hOmega$. 
\smallskip

{\bf Proof:} We decompose $\G=G\cup K$, where $G$ and $K$ are defined by
$G=\{g\in\G : \sigma(g)=1\}$ and $K=\{h\in\G : \sigma(h)=-1\}$. It is then
immediate that $G$ is a (normal) subgroup of $\G$. By assumption
$\hOmega\in K$, and if $h\in K$, then $\hOmega^{-1} h \in G$, as 
$\sigma(\hOmega^{-1} h)=\sigma(\hOmega)^{-1}\cdot\sigma(h)=1$.  
Since $h=\hOmega \hOmega^{-1} h$, if follows that 
$h\in\hOmega G$. Conversely it is clear that $\hOmega G\subset K$,
and we have thus shown that $\G=G\cup\hOmega G$. We can use a similar
argument to show that $\G=G\cup G\hOmega$, and it therefore follows that 
$\hOmega G = G\hOmega$.
\smallskip

To truncate the string theory to the invariant subspace of $\G$, we
truncate it first to the invariant subspace of $G$ as before. Then, since 
$G\hOmega = \hOmega G$, the operator $\hOmega$ preserves $\A^G$
and we can truncate to the $\hOmega$-invariant subspace of $\A^G$ as
explained in \cite{GZ1}; the resulting space is precisely the $\G$
invariant subspace of $\A$. In particular, this demonstrates that the 
invariant subspace of an arbitrary orientifold group can be obtained as  
the truncation of a theory with a basic algebraic structure by a 
{\it single} order reversing 
element.\footnote{In \cite{GZ1} we discussed a simultaneous truncation
defined by the twist operators $\omega_+$ and $\omega_-$ which are
associated to the matrices $J_+$ and $J_-$. In the present language the
orientifold group is $\G=G\cup\omega_{+} G$, where $G=\{\bbbone , J_-\}$,
and $J_{-}$ denotes the inner automorphisms generated by $J_-$. The truncation
by $G$ restricts $\M_{2n}(\Cop)$ to the subalgebra
$\M_{n}(\Cop)\oplus\M_{n}(\Cop)$, and the truncation of this subtheory by 
$\omega_{+}$ leads to the $u(n)$ theory as described at the end of
section~5.6.} 
\smallskip

Orientifold group truncations can be applied to the original open string
theory itself; the main interest, however, lies in the situation where it
is applied to a suitable tensor theory. Suppose then that we are
considering a tensor theory $\A\o\a$, and suppose that an orientifold group
$\H=H\cup\hat\omega H$ acts on the algebra $\a$ so that the analogues of
(i), (ii), (iii'), (iv) hold for the even bilinear form, the products and
the star conjugation defined on $\a$. If the original open string theory
$\A$ has an orientifold group $\G=G\cup\hOmega G$, the tensor 
theory is readily seen to have the orientifold group 
$(G\times H) \cup (\hOmega G \times \hat\omega H)$,\footnote{This
is the direct product $\G\times \H$ in the category whose objects are
groups with $\Zop_2$-homomorphisms, together with the appropriate
notion of morphisms.} where
$(g,h)\in G\times H$ acts on  $(A\o a) \in \A\o\a$ as $g(A)\o h(a)$, and
likewise for 
$(\hOmega g,\hat\omega h)\in(\hOmega G\times\hat\omega H)$.
If $G$ and $H$ have a common subgroup $K$, the tensor theory has the
orientifold subgroup $(K\times K) \cup (\hOmega K \times \hat\omega K)$, and
the theory can be truncated with respect to the diagonal orientifold
subgroup. 
\smallskip

As an example, let us consider an open string field theory in 10
dimensions defined by a NS sector without GSO projection. This theory has a
natural $G= \Zop_2$ symmetry, where the non-trivial generator acts as
$(-1)^{F_w}$, and $F_w$ is the worldsheet fermion number operator 
($(-)^{F_w}$ acts as $(-1)$ on the open string tachyon, as $(+1)$ on the
massless vectors, {\it etc.}). In this case, the invariant subspace $\A^G$
is the ordinary GSO projected NS open string (which we shall denote by
NS(GSO)), and the eigenspace corresponding to the non-trivial
representation of $\Zop_2$ is the NS open string with the opposite GSO
projection (which we shall denote by NS($-$GSO)).  

We take $\a=gl(n+m,\Cop)$, and we define a $H= \Zop_2$ action on $\a$ by
defining the non-trivial generator $\tau$ to act as 
\be
\label{tau}
\tau \left(
\begin{array}{cc}
A & B \\
C & D
\end{array}
\right) = 
\left(
\begin{array}{cc}
A & - B \\
- C & D
\end{array}
\right)
\,,
\ee
where $A, B, C$, and $D$ are 
$n\times n$, $n\times m$, $m\times n$, and $m\times m$ matrices,
respectively. Actually $\tau$ is an inner automorphism of the matrix
algebra, as $\tau (M) = L M L^{-1}$, where 
$L = \hbox{diag}(\bbbone_n, -\bbbone_m)$.
This implies directly that (i) and (ii) hold. It is also easy to see that 
$\tau$ commutes with the transposition ($\omega (A) = A^t$) and with
the hermitian conjugation (star operation).   

We can now truncate the tensor theory with respect to the diagonal $\Zop_2$
whose nontrivial element is $(-)^{F_w} \o \, \tau$, and we find that the
resulting theory is given by
\be
\mbox{NS(GSO)} \o 
\left(
\begin{array}{cc}
A & 0 \\
0 & B
\end{array}
\right) 
\bigoplus
\mbox{NS} ( - \mbox{GSO} )  \o 
\left(
\begin{array}{cc}
0 & C \\
D & 0
\end{array}
\right)\,.
\ee
We can then consider the $\bOmega=+1$ subspace, and the low lying states of
this theory consist of gauge bosons in the adjoint representation of 
$SO(n)\times SO(m)$, together with tachyons in the bi-fundamental
representation $({\bf n},{\bf m})$. The case $n=m=32$ is precisely the open
string sector of the $SO(32)\times SO(32)$ discussed in \cite{BSag,BG}. 
\smallskip

A more sophisticated example is provided by the theory which consists of
an unprojected NS and an unprojected R sector, and which has a 
$\Zop_2\times \Zop_2$ symmetry generated by $(-)^{F_s}$ and $(-)^{F_w}$,
where $F_s$ and $F_w$ are the spacetime and worldsheet fermion number
operator, respectively. The space of states then decomposes into four
sectors which are labeled by $[i_s, i_w]$, where $i_s$ and $i_w$ denote the
eigenvalues of $(-)^{F_s}$ and $(-)^{F_w}$, respectively. Let
$\a=\M_n(\Cop)$ be the internal algebra, and consider the block diagonal
matrices $L_1 = \hbox{diag}(\bbbone_p, -\bbbone_q, 
\bbbone_{\bar p}, -\bbbone_{\bar q} )$, and
$L_2= \hbox{diag}(\bbbone_{p+q}, -\bbbone_{\bar p + \bar q} )$, where
$n=p+q+\bar{p}+\bar{q}$. We introduce a $\Zop_2\times\Zop_2$ action on $\a$
which is generated by the commuting operators $\tau_1$ and $\tau_2$, where
$\tau_i (M) = L_i M L_i^{-1}$, $i=1,2$. The matrix algebra $\a$ then
decomposes into four sectors which are labeled by $(\tau_1, \tau_2)$, where
$\tau_i$ is the eigenvalue of the operator $\tau_i$. The truncation to
the diagonal subgroup whose non-trivial elements are $(-)^{F_s}\tau_1$ and 
$(-)^{F_w}\tau_2$, leads to a theory defined by
$\sum_{i,j=\pm}[i,j]\o (i,j)$. If we truncate the theory further to the
twist invariant sector, we obtain the gauge group
$SO(p)\hskip-2pt\times \hskip-2pt SO(q)\hskip-2pt \times \hskip-2pt 
SO(\bar{p})\hskip-2pt \times \hskip-2pt SO(\bar q)$ and the other
lowest lying states are tachyons in the bi-fundamentals
$({\bf p},\bar{\bf p})$ and $({\bf q},\bar{\bf q})$, massless fermions of
one chirality in $({\bf p},{\bf q})$ and $(\bar{\bf p}, \bar{\bf q})$, and
massless fermions of the opposite chirality in $({\bf p},\bar{\bf q})$ and 
$(\bar{\bf p}, {\bf q})$. The 
$SO(32-n)\hskip-2pt\times\hskip-2pt SO(n)\hskip-2pt \times\hskip-2pt 
SO(32-n) \hskip-2pt\times \hskip-2pt SO(n)$ theories considered in
\cite{BSag,BG} are of this type.

\section{Open string vector bundles}
\setcounter{equation}{0}

So far we have only discussed tensor constructions where the resulting
space of states is the full product space. We shall now consider tensor
constructions which are analogous to the tensor product of vector
bundles. 

\subsection{Bundle constructions and their tensoring}  

A string theory which is described in terms of a homotopy 
associative algebra $\A$ has the structure of a {\it vector bundle} over a
semigroup $S$ with zero if the following conditions are met
\begin{list}{(\roman{enumi})}{\usecounter{enumi}}
\item $\A$ has the decomposition
$\A = \bigoplus_{s\in S} \A_s$, 
where for $A_i\in\A_{s_i}$, $i=1,\ldots, n$ and $n\in\Nop$
\be
\label{products}
b_n(A_1, \ldots, A_n) \in \A_s \, , \quad \hbox{where}\quad  s= s_1\cdot s_2
\cdot\, \cdots \, \cdot s_n \,.
\ee
\item There exists an involution $\ast:S\rightarrow S$ so that 
$A^\ast\in\A_{s^\ast}$ for all $A\in\A_s$.
\item There exists an involution $c:S\rightarrow S$, so that for $A\in\A_s$
and $B\in\A_t$, $\langle A,B\rangle = 0$ if  $t\not= c(s)$.
\end{list}
Sometimes we may also require that there exists an involution
$\omega : S\rightarrow S$, so that for $A\in\A_s$, 
$\Omega(A)\in\A_{\omega(s)}$. 

We may assume without loss of generality that there exists only one 
$s\in S$, for which $\A_s=\empty$, and we shall write $A_0=\empty$. 
(For otherwise, we can replace $S$ by a semigroup where all $s\in S$ for
which $\A_s=\empty$ are identified.) With this definition, it is clear that
the involutions $c$, $\ast$ and $\omega$ all map $0\mapsto 0$. 
It is then natural to consider decompositions that are {\it minimal} in the
sense that if $s_1 s_2 \ne 0$, there exists $A_i\in\A_{s_i}, i=1,2$ so that
$b_2(A_1,A_2)\ne 0$.
\medskip

Let us now consider the situation, where the string field theory $\A$ has
the structure of a vector bundle over $S$. Let us furthermore assume that
the algebra $\a$ that we wish to tensor to $\A$ satisfies the
analogous properties to (i)-(iii) with respect to the same semigroup
$S$ and the same involutions $c$ and $\ast$. Then we can define the
fiberwise tensor product by the following construction. The ordinary tensor
product $\A\o\a$ contains the subspace
\be
\label{fiberpro}
\A\o\a = \bigoplus_{s,t\in S} \A_s \o \a_t \supset
\bigoplus_{s\in S} \A_s \o \a_s =:
\A\o_S \a \,.
\ee
This subspace has the structure of a homotopy-associative algebra, as (i)
guarantees that $\A\o_S \a$ is closed under the various
products. Furthermore, because of (iii), the restriction of the bilinear
form to $\A\o_S \a$ is invertible, and the star operation is well-defined
on $\A\o_S \a$ because of (ii). It thus follows that $\A\o_S \a$ has the
basic algebraic structure which guarantees that the associated string field
action satisfies the classical master equation. Furthermore, it is also
clear from (\ref{fiberpro}) that $\A\o_S \a$ has
the structure of a vector bundle over $S$.
\smallskip

If $S$ is an abelian group, we can define an action of $\widehat{S}$,
the dual group of $S$, on $\A$ by $\rho (\psi_s) = \rho(s) \psi_s$, where
$\psi_s\in\A_s$ and $\rho\in\widehat{S}$. Then the decomposition
of $\A$ can be interpreted as the decomposition into
irreducible representations of $\widehat{S}$, and the fiberwise tensor
product is just the truncation of the ordinary tensor product to the invariant
subspace under the action of $\widehat{S}$. (Here $\rho\in\widehat{S}$ acts
on $(A\o a)\in\A_s\o\a_t$ as $\rho(A\o a) = \rho(s) \rho^{-1}(t) (A\o a)$.)
In particular, every projection of a tensor theory by an abelian group can
be interpreted as a bundle tensor product; all presently known projection
constructions are of this type.

If $S$ is a group, and if the fibration of $\A$ is trivial in the sense
that $\A_s\cong\A_e$ for every $s\in S$ (where $e$ is the group identity in
$S$), then the fiberwise tensor product $\A\o_S \a$ is equivalent to 
the ordinary tensor product $\A_e \o \a$.

\subsection{Brandt semigroups}

Before we exhibit some more explicit examples we want to show that
the algebraic structure of the string field theory implies (under certain
conditions to be explained below) that $S$ is actually a union of
{\it Brandt semigroups} \cite{CP} and trivial semigroups. In order to
prepare for the discussion of this theorem, we want to introduce in this
subsection some of the relevant notions.

A Brandt semigroup is a semigroup with zero, for which  
\begin{itemize}
\item[(1)] for every element $s\in S$, $s\ne 0$, there exist unique
elements $e,f$ and $s'$ so that $es=s$, $sf=s$, and $s's=f$. $e$, $f$ and
$s'$ are called left-identity, right-identity, and inverse,
respectively.\footnote{It follows directly from the uniqueness of the 
identities and the inverse that $ee=e$, $ff=f$, $ss'=e$ and
$s'e=s'$, $fs'=s'$ \cite{CP}.}
\item[(2)] If $e$ and $f$ are non-zero and satisfy $ee=e$ and $ff=f$, then
there exists $s\in S$ so that $esf=s$.
\end{itemize}

If $S$ is a semigroup satisfying (1) then $S$ can be written as a direct
union of Brandt semigroups $S_i$, $i=1,\ldots, n$, where 
$S_i \cap S_j = \{0\}$ for all $i\ne j$, and 
$s_i s_j = 0$ for $s_i\in S_i$, $s_j\in S_j$, and $i\ne j$. 

Indeed, let $e_\alpha$, $\alpha=1, \ldots, N$ denote 
the non-zero idempotents in
$S$, {\it i.e.} the elements $e\ne 0$ which satisfy $ee=e$. Because of (1)
we can decompose $S$ as a disjoint union
\be
S= \bigcup_{(\alpha,\beta)} S_{\alpha,\beta} \cup\{0\}\,,
\ee
where $S_{\alpha,\beta}$ consists of those non-zero elements in $S$ for
which  $e_\alpha$ is the left-identity and $e_\beta$ the right-identity. 
We introduce a relation on the set $\{1, \ldots, N\}$ by defining
$\alpha\sim \beta$ if $S_{\alpha,\beta}\ne\empty$ (so that when 
$\alpha \sim\beta$, $e_\alpha$ and $e_\beta$ satisfy (2)). This relation 
is reflexive as $e_\alpha\in S_{\alpha,\alpha}$, it is symmetric, as 
$s\in S_{\alpha,\beta}$ implies that $s'\in  S_{\beta,\alpha}$, and it is
transitive because $ab\not=0$ if and only if the right identity of $a$ 
equals the left identity of $b$ \cite[Lemma 3.8]{CP}. 
The relation is therefore an equivalence
relation, and we can decompose $\{1, \ldots, N\}$ into the various
equivalence classes which we denote by  
$\Delta_i$, $i=1,\ldots, n \leq N$. 
We can then decompose $S$ as a direct union
\be
S= \bigcup_{i=1}^{n} S_i\,, \qquad \mbox{where} \qquad
S_i = \bigcup_{\alpha_i,\beta_i\in\Delta_i}
S_{\alpha_i,\beta_i} \cup\{0\}\,,
\ee
where each $S_i$ is a Brandt semigroup and the decomposition satisfies 
the above conditions.

\subsection{The bundle structure of open string theory}

We can now classify the possible base spaces of vector bundles which
describe string theories; this result relies crucially on the cyclicity
properties of the bilinear form.
\smallskip

{\bf Theorem:} Let $\A$ be a string theory which has a 
minimal decomposition as a
vector bundle over a semigroup $S$ with zero. Furthermore, let us
assume that every $\A_s$ for $s\ne 0$ contains non-trivial physical
states, and that $c_0$ maps each sector to itself.
Then $S$ is a direct union of Brandt semigroups and trivial semigroups
(semigroups  of the form $T_j=\{e_j,0\}$ where $e_j e_j =0$.).  
\medskip

{\bf Proof:} First we prove that 
$c=\ast$ in $S$. For each $s\ne 0$ we consider a non-trivial physical state
$\psi_s\in\A_s$. Because of the positivity of the inner product it then
follows that $\langle \psi^\ast_s, c_0 \psi_s \rangle \ne 0$. 
Since $c_0 \psi_s \in \A_s$, (iii) (sect.~4.1) implies that
$\psi^\ast_s\in\A_{c(s)}$, and since $\psi^\ast_s\in \A_{s^*}$ ((ii)
sect.~4.1), it follows that $c(s)=s^\ast$ for all $s\ne 0$. On the other
hand $c(0)=0^\ast= 0$, and we have shown that $c=\ast$ on $S$.
\smallskip

Next, let $e_j$, $j=1,\ldots, p$ denote the trivial elements of $S$, 
{\it i.e.} the elements which satisfy $e_j s=se_j=0$ for every $s\in S$. 
$S$ is then a direct union of the trivial semigroups $T_j=\{e_j,0\}$ and
$S_0$, where for every non-zero element $s\in S_0$, there exists
$\hat{s}\in S_0$ so that either $s\hat{s}\ne 0$ or $\hat{s} s\ne 0$. We
want to show that $S_0$ is a direct union of Brandt semigroups.

Let $s\in S_0$ be non-zero, and let us assume that $s\hat{s}\ne 0$ for a
suitable $\hat{s}\in S_0$. (The case where $\hat{s} s\ne 0$ can be
analyzed analogously.) Because of the minimality of the decomposition we
can find $A_s\in\A_s$ and $A_{\hat{s}}\in\A_{\hat{s}}$ so that
$b_2(A_s,A_{\hat{s}}) \equiv A_sA_{\hat s}\ne 0$. 
As the bilinear form is non-degenerate, there 
exists $A_{s\hat{s}}\in\A_{(s \hat{s})}$ so that   
$\langle A_{s\hat{s}}^\ast, A_sA_{\hat s} \rangle \ne 0.$
Using the hermitian conjugation of the bilinear form together with the
action of the star conjugation on products, this implies that 
$\langle A_{s\hat{s}}, A^\ast_{\hat s}A^\ast_s \rangle \ne 0$.
Because of the cyclicity of the bilinear form we then have
$\langle A^\ast_{\hat s}, A^\ast_s A_{s\hat s}) \rangle \ne 0$,
and therefore $A^\ast_s A_{s\hat s}\in\A_{\hat{s}}$. We can thus
conclude that $s^\ast (s \hat{s}) = \hat{s}$; this implies, in particular,
that $s^\ast s\ne 0$. 

This identity holds whenever for a given $s\not=0$, $s\hat s\not=0$. We
can therefore apply the identity with $(s,\hat{s})$ being replaced by 
$(s^\ast, s)$, and we thus find that $s s^\ast s=s$.
This implies that $s s^\ast$ is a left-identity, $s^\ast s$ a right-identity,
and $s^\ast$ an inverse of $s$. In order to show (1) in the definition of
the Brandt semigroup, it therefore only remains to prove that the two 
identities and the inverse are unique. Suppose then that $e$ is a
left-identity of $s\ne 0$. By the minimality of the decomposition we can
then find  $A_e\in\A_e$ and $A_s,A'_s\in\A_s$ so that  
$\langle {A'}_s^\ast, A_eA_s\rangle \ne 0$.
Using the same arguments as above, this implies that 
$\langle A_e^\ast, A'_sA_s^\ast \rangle \ne 0$,
and thus that $e=ss^\ast$. The arguments for the uniqueness of the
right-identity and the inverse are similar. This proves (1).
By the arguments of the previous section, it then follows that
$S_0$ is a direct union of Brandt semigroups. This completes the proof.  
\bigskip

Essentially the same proof can also be used to show that for every
fibration of a full matrix algebra over a semigroup $S$, $S$ has to
be a direct union of Brandt semigroups. (In this case, it is clear that
there are no trivial semigroups.)
\smallskip

The characterization of $S$ as a direct union of Brandt semigroups and
trivial semigroups is useful, as Brandt semigroups are completely
classified \cite{CP}. (This is a special case of Rees' Theorem.)
Indeed, every Brandt semigroup is uniquely characterized by a group $G$,
and a positive integer $n$. The non-zero elements of the corresponding
Brandt semigroup are described by triples $(g,[i,j])$, where $g\in G$,
$i,j\in\{1,\ldots,n\}$, and the semigroup product is defined by  
\be
(g,[i,j]) (h,[k,l]) =  \left\{
\begin{array}{cl}
(gh,[i,l]) & \mbox{if $j=k$,} \\
0 & \mbox{otherwise.}
\end{array}
\right.
\ee
This structure is realized in a configuration of $n$ types of branes. 
In this case the theory decomposes into sectors $[ij]$ 
\be
\label{breakup}
\A = \bigoplus_{i,j=1}^n\,\, [\, ij\,] \,,
\ee
where $[ij]$ contains all strings that begin on a brane of type $i$, and
end on a brane of type $j$. In addition, each sector $[ij]$ decomposes 
into subsectors $[g,[ij]]$ that are labeled by elements $g\in G$ so
that $[ij]=\cup_{g\in G} [g,[ij]]$. In the following we shall limit ourselves
to the case where $G=1$, but interesting constructions with $G\ne 1$ may
exist. 

By construction, the product of sectors satisfies
$[ik]\,\times[lj]=[ij]\,\delta_{kl}$, and the twist operator $\Omega$ maps
$[ij]$ to $[ji]$, as this operation changes the orientation of the open
strings. The same also applies to the star conjugation, and we have
$\omega=c=*$ on the semigroup $S$.

\section{Multi-brane tensor theories}
\setcounter{equation}{0}

We shall now consider the fiberwise tensor product of a string field theory of
the form (\ref{breakup}) with a matrix algebra $\a$. In order to
proceed we have to decompose $\a=\M_N (\Cop)$ as a vector bundle over the
same semigroup $S$. This is done by choosing a partition of $N$
into $n$ positive numbers $r_1, \ldots, r_n$ (so that $\sum_i r_i=N$), and
by decomposing every matrix $a\in\M_N(\Cop)$ into $n^2$ blocks, where we
denote by $a_{(ij)}$ the $(i,j)$ block which has $r_i$ rows and $r_j$
columns. We then decompose the matrix algebra $\M_N (\Cop)$ as
\be 
\M_N (\Cop) = \bigoplus_{i,j=1}^n \, ( ij) \,, 
\ee 
where $(ij)$ is the subspace of those matrices for which all the
non-zero entries lie in the $(i,j)$ block. It is then obvious that 
$(ik)\,\times(l j)=(ij)\,\delta_{kl}$. Furthermore, the bilinear form
(defined by the trace) couples $(ij)$ to $(ji)$, and the star operation
(hermitian conjugation) maps $(ij)$ to $(ji)$. Later on, we shall also
define a twist operator $\omega$ which maps $(ij)$ to $(ji)$.

The fiberwise tensor product gives then 
\be
\label{tenf}
\A\o_S\a= \bigoplus_{i,j=1}^n \, \Bigl( [ij]\, \otimes \, (ij) \Bigr) \,.  
\ee 
This can be interpreted as a configuration, where we have $r_i$ branes of
type $i$, where $i=1,\ldots, n$. The open strings that begin and end on the
branes of type $i$ (the $[ii]$ sector) carry the gauge group $U(r_i)$, and
the total  gauge group is $U(r_1)\times \cdots \times U(r_n)$. On the other
hand, the open strings that begin on a brane of type $i$, and end on a
brane of type $j$, $i\ne j$ (the $[ij]$ sector) transform in the
bi-fundamental  $({\bf r}_i,\bar { {\bf r}}_j)$ of the gauge (sub)group  
$U(r_i)\times U(r_j)$. 
\bigskip

In bosonic string theory $\Omega^2=+1$, but this need not be the case for
fermionic strings, where there exist sectors for which $\Omega^2=-1$
\cite{GimPol}. Let us assume that the decomposition of $\A$ in
(\ref{breakup}) is such that all states in a given sector $[ij]$ have the
same $\Omega^2$ eigenvalue which we denote by $\Omega^2_{ij}$. 
As $\Omega^2$ is an automorphism of the algebra $\A$, it follows that 
\be
\label{osq}
\Omega^2_{ij} = \Omega^2_{ik} \Omega^2_{kj}\,, 
\ee
where $i,j$ and $k$ are arbitrary. We can choose $i=j=k$, and (\ref{osq})
then implies that $\Omega^2_{ii} =1$, for all $i$. 
Next we choose $p_1\ne 0$ to be arbitrary, and we define $n-1$ constants
$p_i\ne 0$, $i=2,\ldots, n$ by $\Omega^2_{j1}= p_j / p_1$. It then follows
from (\ref{osq}) that the general element is  
\be
\label{param}
\Omega^2_{ij} = {p_i \over p_j} \,. 
\ee 
On the other hand, it is easy to see that (\ref{param}) defines a solution
of (\ref{osq}) for every choice of $p_i\ne 0$, $i=1,\ldots, n$. Two choices
of $p_i$ which are related by $\hat{p}_i = \mu p_i$ define the same
solution for $\Omega^2_{ij}$.
We shall be mainly interested in the case where $\Omega^l = +1$ for some
integer $l$, and in particular, in the case where $l=4$; in this case we
can choose $p_i$ so that $p_i=\pm 1$ for all $i=1,\ldots, n$.
\medskip

We want to define the twist operator $\omega$ on $\a$ in such a way that 
${\bf \Omega} = \Omega\o \omega$ satisfies $\bOmega^2=+1$, so that 
we can project the theory to eigenstates of $\bOmega$ of eigenvalue
$+1$.\footnote{If $\bOmega^2\ne +1$, then some open string sectors would be
projected out altogether. While this is consistent classically, the theory
would presumably be inconsistent upon the inclusion of closed strings.} 
This implies that $\omega^2$ must take the same eigenvalue 
$\omega^2_{ij}$ for all states in a given sector $(ij)$, and that
$\omega^2_{ij} = 1/ \Omega^2_{ij}$. Because of (\ref{param}), it then
follows that  
\be
\label{consq}
\omega^2 (a) = M^{-1} \, a \, M\,, \quad
M_{(ij)} =p_i\delta_{ij} \, \bbbone_{r_i\times r_i}\,, 
\ee
where, as indicated, the matrix $M$ is formed of diagonal blocks, each of
which is a matrix proportional to the identity. As reviewed in section~2,
$\omega$ is of the form
\be
\omega(a) = K a^t K^{-1} \,,
\ee
and since $\omega$ maps $(ij)$ to $(ji)$, $K$ must be block diagonal,
{\it i.e.} $K_{(ij)} = \delta_{ij} K_i$, where
$K_i$ is an $r_i\times r_i$ matrix. Using Schur's lemma, it follows from
(\ref{consq}) that $K^t K^{-1} = \lambda  M$, where $\lambda$ is some
constant. This implies that
$K_i^t = \lambda \, p_i \, K_i$
and consistency requires that $(\lambda p_i)^2=1$. For the case where
$\Omega^4=1$, $p_i=\pm 1$, this implies $\lambda^2=+1$, and by replacing
$p_i$ by $-p_i$ if necessary, we may assume without loss of generality that
$\lambda=+1$. In this case $K_i^t = p_i K_i$, and the matrix $K_i$
is symmetric for $p_i=+1$, and antisymmetric for $p_i=-1$. (As
$K$ is nondegenerate, $r_i$ must be even whenever $p_i=-1$.) By a change
of basis, we can bring $K$ to canonical form
\be
K_{(ij)} = \delta_{ij} J_{p_i}^{(r_i)} \,,
\ee 
where $J_{+1}^{(r)}$ is the unit $r\times r$ matrix, and $J_{-1}^{(r)}$ is
the symplectic $r\times r$ matrix (with $r$ even). Finally, $\omega$ is
then given as
\be 
(\omega (a))_{(ij)} = p_j \cdot J_{p_i} \, (a_{(ji)})^t \,
J_{p_j} \, .  
\ee 
It can be checked that the action of $\omega$ commutes with that of the
star involution; this requires that $K^\dagger K$ is proportional to the
identity, which is readily verified.  In the diagonal blocks the above
action of $\omega$ gives 
\be (\omega (a))_{(ii)} = p_i \cdot J_{p_i} \, (a_{(ii)})^t \, J_{p_i} \,.
\ee 
This leads to a gauge group $SO(r_i)$ for each $p_i=+1$, and a gauge group
$USp(r_j)$ for each $p_j=-1$.  This is in accord with the result of
\cite{GimPol} for the case of a system of 5- and 9-branes.

\section*{Acknowledgments}

\noindent M.R.G. is supported by a 
NATO-Fellowship and in part by NSF grant PHY-92-18167.
B.Z. is supported in part by D.O.E.
contract DE-FC02-94ER40818, and a fellowship of the John Simon Guggenheim
Memorial Foundation.

\end{document}